%% file: Main.tex
\documentclass[twocolumn]{NobArticle}
\runninghead{Assessing the ROI of CTI}

\usepackage{darkmode}
\usepackage{orcidlink}
\usepackage{pgfplots}
\pgfplotsset{compat=1.18}

\newcolumntype{L}[1]{>{\raggedright\let\newline\\\arraybackslash\hspace{0pt}}m{#1}}

\title{Assessing the ROI of Cyber Threat Intelligence: An Operational and Financial Evaluation Framework}

\author{%
    \begin{minipage}[t]{0.48\linewidth}
        \raggedright
        Matteo Strada\,\orcidlink{0009-0003-3197-2662}\\
        Computer Science Department, University of Milan\\
        \texttt{matteo.strada@studenti.unimi.it}\\
    \end{minipage}%
    \hfill%
    \begin{minipage}[t]{0.48\linewidth}
        \raggedright
        Stelvio Cimato\,\orcidlink{0000-0003-1737-6218}\\
        Computer Science Department, University of Milan\\
        \texttt{stelvio.cimato@unimi.it}
    \end{minipage}%
}

\date{}

\fancypagestyle{plain}{%
  \fancyhf{}%
  \fancyfoot[C]{\footnotesize Preprint version. Under review.}%
  \fancyfoot[RO,LE]{\small\textbf{\thepage}}%
  \fancyfoot[LO,RE]{\footnotesize\footertext}%
}
\makeatletter
\patchcmd{\maketitle}{\thispagestyle{empty}}{\thispagestyle{plain}}{}{}
\makeatother

\renewcommand{\maketitlehookd}{%
\vspace*{-1.3\baselineskip}
\begin{abstract}
    \noindent 
    Quantifying the Return on Investment (ROI) of Cyber Threat Intelligence (CTI) poses a measurement problem: successful prevention produces non-events that leave no observable financial signal, which makes CTI spending resistant to traditional cost-benefit analysis. CTI's indirect contribution through downstream controls further complicates causal attribution and the investment boundary. We develop a framework with two main contributions: (i) the Threat Intelligence Effectiveness Index (TIEI), a weighted geometric maturity measure spanning intelligence quality, enrichment, integration, and operational impact that penalizes weak links; and (ii) a breakeven-first financial method that treats the annual probability of a scope-matched material event and CTI-attributable mitigation as unknowns and characterizes the combinations required for positive ROI. The financial boundary includes core CTI ownership and the incremental downstream costs required to operationalize intelligence. Applied to illustrative finance and healthcare scenarios, the method demonstrates how organizations can derive scope-matched breakeven requirements without treating broad sector prevalence as an event probability. A TIEI-conditioned PERT simulation illustrates how uncertainty can be propagated after an organization supplies a defensible event probability, while operational and qualitative indicators support attribution where avoided events cannot be observed. By linking operational maturity to an auditable financial boundary, the framework provides a reproducible basis for CTI investment decisions.

    \medskip

    \small{\textbf{Index Terms:} Cyber Threat Intelligence (CTI), Return on Investment (ROI), Return on Security Investment (RoSI), Threat Intelligence Effectiveness Index (TIEI), Cybersecurity Economics, Risk Quantification, FAIR model}
\end{abstract}
}

\begin{document}

\small
\maketitle

\input{Sections/01-Introduction}
\input{Sections/02-Methodologies}
\input{Sections/03-UseCases}
\input{Sections/04-Challenges}
\input{Sections/05-Conclusion}

\FloatBarrier

\input{Sections/06-Appendix}

\printbibliography

\end{document}

%% file: Sections/01-Introduction.tex
\section{Introduction}
\label{ch:introduction}
\subsection{Cyber Threat Intelligence: Definition and Scope}

Cyber Threat Intelligence (CTI) is defined as evidence-based knowledge, including context, mechanisms, indicators, and actionable advice, concerning existing or emerging threats to organizational assets \cite{NISTCTIDefinition}.
The fundamental purpose of CTI is to transform high volumes of raw data into actionable insights that inform and improve cybersecurity decisions. This systematic process enables an organization to mature its security posture, transitioning from a reactive stance, where teams respond to incidents as they occur, to a proactive one, where attacks are anticipated and preemptively mitigated \cite{NISTCTIDefinition}. 

CTI can be stratified into three distinct levels, each serving a different audience and purpose within an organization. \emph{Strategic intelligence} is high-level, non-technical intelligence designed for executive leadership and the board of directors. It provides a broad overview of the global threat landscape, geopolitical trends, and industry-specific risks. Its primary function is to inform long-term cybersecurity investment, risk management strategy, and the alignment of security with overall business objectives \cite{RecordedFutureLevels}.

\emph{Operational intelligence} offers more technical detail regarding the actors, motivations, methods, and objectives associated with specific cyberattack campaigns. It focuses on the tactics, techniques, and procedures (TTPs) of threat actors. The primary consumers are security leaders, threat hunting teams, and incident responders, who use this intelligence to understand adversary behavior and prepare specific defenses \cite{RecordedFutureLevels}.

\emph{Tactical intelligence} is the most immediate and technical form of CTI, consisting primarily of machine-readable Indicators of Compromise (IOCs) such as malicious IP addresses, file hashes, and phishing domains. It is consumed by Security Operations Center (SOC) analysts and automated security tools (e.g., SIEMs and firewalls) for real-time threat detection and blocking. While highly actionable, tactical IOCs have a short lifespan because adversaries frequently change their infrastructure \cite{bianco2014pyramid, RecordedFutureLevels}.

A failure to recognize and measure the value delivered at all three levels may contribute to incomplete CTI ROI calculations. A comprehensive ROI model must account for the distinct forms of value provided to different stakeholders, from the operational efficiency gained by a SOC analyst using tactical feeds to the strategic risk reduction achieved by a board using high-level threat landscape reports.

\subsection{The Measurement Gap in CTI ROI}
As cybersecurity investments continue to grow, organizations face increasing pressure to justify their financial investments in this domain. Rather than simply asserting that security will improve, there is growing pressure to demonstrate the value of these investments through clear, measurable data.

This demand for accountability presents a significant challenge for CTI programs. The 2025 SANS CTI Survey revealed that a primary struggle for CTI teams is the difficulty in proving ROI and securing executive support \cite{SANS2025Survey}. This finding is supported by research from ESG, which found that 71\% of security professionals report difficulty in measuring the ROI of their CTI program \cite{ESG2023}. Without a clear and defensible model for quantifying its contribution to risk reduction and business value creation, a CTI program risks being viewed as a cost center rather than a strategic asset. 

This perception can lead to underfunding, resource misallocation, and deprioritization in favor of other security initiatives that offer more easily measured returns. Therefore, establishing a robust framework for calculating and communicating CTI ROI is an important factor in sustaining CTI programs and demonstrating their organizational value \cite{SANS2025Survey}.

\subsection{Limitations of Traditional Security Investment Models}
\label{sec:economics}
Historically, cybersecurity investments have been evaluated using traditional optimal-spending models, with the most prominent being the \textit{Gordon--Loeb (GL) model} \cite{gordon_loeb_2004}. The GL framework evaluates the financial return of implementing a technical control to protect a specific asset. It considers an information asset that would incur a one-time loss $L$ if a breach succeeds and whose baseline vulnerability (probability of breach without further protection) is $v\in(0,1)$. In Gordon and Loeb's notation, $S(z,v)$ is the residual breach probability after investing $z$ in a security control. Their general assumptions require $S(0,v)=v$, $S(z,0)=0$, a decreasing and strictly convex response to investment for $v>0$, and $S(z,v)\rightarrow0$ as $z\rightarrow\infty$. The optimal spend is
\[
  z^{*}=\arg\min_{z\ge0}\bigl\{\,z+LS(z,v)\bigr\}\,.
\]
For the two functional families analyzed explicitly by Gordon and Loeb, the optimum satisfies
\[
  z^{*}\;<\;\frac{1}{e}\,vL\;\approx\;0.368\,vL.
\]
This $1/e$ rule is therefore not a universal consequence of the general assumptions. Willemson constructs admissible security-breach functions satisfying Gordon and Loeb's original smoothness conditions for which optimal investment approaches $vL/2$. He obtains optima arbitrarily close to the full expected loss $vL$ only after relaxing the requirement that the second derivative be continuous \cite{willemson2006on}. The unconditional result under the original general assumptions is the weaker bound $z^{*}<vL$.

However, applying traditional models like GL to CTI reveals a fundamental category error. GL is designed to evaluate independent technical controls. CTI, conversely, is an informational asset. CTI does not stop a breach on its own; rather, it acts as a force multiplier that optimizes the efficacy of secondary controls (e.g., firewalls, EDR, and human analysts). In practice, CTI influences risk outcomes along multiple dimensions: it can lower the probability that an intrusion remains undetected, shorten attacker dwell time, thereby reducing the conditional loss given breach, and increase the attacker's cost structure. Empirical studies confirm that intelligence-driven SOC workflows cut mean time-to-detect and mean time-to-respond \cite{saeed_2023,mandiant_2024}. Because traditional economic models struggle to quantify this indirect leverage effect, this paper shifts from traditional control-based optimization toward a probabilistic risk-reduction model.

This paper makes two contributions. First, it proposes the Threat Intelligence Effectiveness Index (TIEI) as a structured internal measure of CTI program effectiveness across four dimensions: intelligence quality, enrichment, integration, and operational impact. Second, it develops a breakeven-first decision framework illustrated through finance and healthcare scenarios. Because neither the organization-specific probability of a material loss event nor the CTI-attributable mitigation fraction is available from broad sector prevalence reports, the primary result is a two-unknown feasibility frontier between those quantities. TIEI-conditioned PERT simulations are retained only as conditional illustrations at explicitly selected event probabilities. The sector values are illustrative rather than predictive and require organization-specific validation before operational use.

%% file: Sections/02-Methodologies.tex
\section{Methodology for Assessing CTI Return on Investment}

An assessment of CTI ROI requires a multi‑layered approach that combines quantitative metrics, qualitative value indicators, and structured financial and operational frameworks. No single data point can capture the full spectrum of value delivered by a mature CTI program; instead, security leaders must assemble a defensible portfolio of empirical indicators addressing efficiency and risk reduction.

A useful way to address the difficulty of measuring cybersecurity return is to treat ROI evidence as a portfolio of observable performance indicators (i.e., proxy metrics) rather than as a single financial data point. In this view, cyber investment value can be inferred from operational measures. These measures are especially relevant when avoided losses are not directly observable, because they provide empirical signals that security capabilities are improving and that the investment is producing measurable operational outcomes~\cite{CyberKPI_ROI}.

\subsection{Quantitative Performance Indicators}
\label{sec:quantitativeKPI}

Quantitative indicators provide tangible, data-driven evidence of CTI's impact on security operations and business outcomes. Where possible, they should be derived from existing telemetry (e.g., SIEM, SOAR, ticketing systems) to minimize additional measurement overhead.

\paragraph{Cost-Avoidance ROI Estimation}
Cost avoidance is one of the main pillars of any cybersecurity ROI calculation. The methodology estimates the potential financial loss of security incidents that were prevented or significantly limited by controls related to CTI. Leading breach cost studies place the average cost of a data breach in the multi‑million dollar range, while sector‑specific research highlights even higher impacts for heavily regulated industries \cite{IBM2025, VerizonDBIR2024}.

A CTI program's contribution can be modeled by mapping its outputs (detections, early warnings, takedowns) to the FAIR (Factor Analysis of Information Risk) framework or similar quantitative risk‑analysis methods. FAIR converts qualitative risk statements into a probabilistic financial range, enabling risk‑adjusted ROI calculations that remain valid even as industry averages evolve \cite{FAIRInstituteWhatIsFAIR, BalbixFAIRModel}.

To move from a conceptual discussion of cost avoidance to a quantifiable model, it is essential to formally define both the total investment and the prospective return. The investment is best captured by the Total Cost of Ownership (TCO) of the CTI program, which includes all direct and indirect expenses.

We can define the total cost of ownership as:
\begin{equation}
\label{eq:CTICost}
  \mathrm{TCO}_{\text{CTI}} = C_{\text{platform}} + C_{\text{feeds}} + C_{\text{personnel}} + C_{\text{infra}} + C_{\text{integration}} + C_{\text{training}}
\end{equation}
where the components represent the annualized costs of CTI platforms and tools, commercial and open-source intelligence feeds, dedicated analyst and engineering personnel, supporting infrastructure, integration and downstream enablement, and ongoing training, respectively. In particular, $C_{\text{integration}}$ includes the incremental cost of detection engineering, EDR/SIEM/SOAR implementation, automation, dependent control changes, and maintenance required to operationalize CTI outputs. Thus, $\mathrm{TCO}_{\text{CTI}}$ is the all-in annual incremental investment required to realize the modeled mitigation, rather than a platform-only price.

With the cost basis established, a direct financial return can be modeled using a cost avoidance formula. This model calculates the value of prevented incidents by factoring in the probability of each incident class, its potential cost, and the effectiveness of CTI in mitigating it.
\begin{equation}
\label{eq:ROIavoidance}
ROI_{\text{avoidance}} = \left( \frac{ \sum_{j=1}^{n} (\lambda_j \times C_j \times M_j) - \mathrm{TCO}_{\text{CTI}} }{ \mathrm{TCO}_{\text{CTI}} } \right) \times 100
\end{equation}

where $\lambda_j$ denotes the organization-specific annualized frequency of material loss event class $j$ under existing controls but before the proposed incremental CTI capability (events per year; for a single rare-event class, $\lambda_j$ approximates its annual probability); $C_j$ is the estimated cost of incident class $j$ if realized; $M_j$ is the CTI mitigation effectiveness factor for incident class $j$, expressed as a value between 0 and 1; $n$ is the total number of modeled incident classes; and $\mathrm{TCO}_{\text{CTI}}$ is the all-in annual incremental investment defined in Equation~\eqref{eq:CTICost}.

The input variables ($\lambda_j$, $C_j$) should be estimated from event-scope-matched internal data, expert elicitation, and probabilistic approaches such as FAIR, with external reports used to inform rather than mechanically replace organization-specific estimates. Annualizing $\lambda_j$ ensures dimensional consistency with annualized $\mathrm{TCO}_{\text{CTI}}$. For a single incident class ($n=1$), the summation reduces to $\lambda C M$; Section \ref{ch:sector_rois} expresses the corresponding rare-event case as a two-unknown feasibility frontier.

\paragraph{Incident‑Related Metrics}
CTI's most visible contribution is the acceleration of the incident‑response lifecycle. By supplying contextual knowledge of threat actors, their tactics, techniques, and procedures (TTPs), and their infrastructure, CTI enables responders to bypass time‑consuming reconnaissance and move directly to containment and eradication. The impact is best expressed through two key performance indicators (KPIs): Mean Time to Detect (MTTD) and Mean Time to Respond (MTTR). Organizations should track trending reductions in these KPIs over consecutive reporting periods and correlate improvements with the introduction of CTI‑driven detections, threat‑hunting hypotheses, or automated playbooks \cite{RecordedFutureLevels}.

A complementary indicator is \emph{attacker dwell time}: the interval between initial compromise and detection. In Mandiant's EMEA investigations for 2024, median dwell time was 32~days for externally notified events and 20~days for internally discovered events, a 12-day difference \cite{MandiantDwellTime2025}. A separate IBM global statistic concerns the broader data-breach lifecycle rather than dwell time: internally detected breaches had a lifecycle 61~days shorter than breaches disclosed by an attacker \cite{IBM2024}. Although these measures are not directly comparable, both findings suggest that stronger internal detection capabilities, supported by timely and operationally integrated CTI, can reduce the period during which an intrusion remains unidentified or unresolved.

\paragraph{Operational Efficiency Metrics}
Security Operations Centers (SOCs) consistently encounter "alert fatigue" driven by high volumes of low‑fidelity alerts \cite{RecordedFutureProductivity2024, DarktraceAlertFatigue}. CTI mitigates this challenge by correlating external threat data with internal context such as asset criticality, exploitability, and business impact; thereby suppressing irrelevant indicators and reducing false positives. Analysts regain time for higher value activities such as proactive hunting and adversary simulation.

Return on investment can be demonstrated by tracking the reduction in alert volume, the percentage of alerts auto‑triaged, and the net analyst hours redeployed to strategic functions. Organizations that integrate CTI enrichment directly into SOAR workflows typically record significantly higher efficiencies than those reliant on manual enrichment or disconnected tools  \cite{RecordedFutureProductivity2024}.
Automated ingestion, normalization, and dissemination of intelligence amplify CTI value. Metrics such as the number of playbooks enriched with CTI, the percentage of enrichment calls completed within service‑level thresholds, and the ratio of automated versus manual escalations provide measurable evidence of scalability and cost savings.

\paragraph{Threat Intelligence Effectiveness Index (TIEI)}
\label{sec:tiei-bridge}

Traditional approaches to evaluating the effectiveness of CTI programs frequently rely on weighted arithmetic means. Although such aggregation methods are computationally straightforward and intuitive, they rest upon the assumption of linearity, implicitly treating each incremental improvement as equally valuable across the performance spectrum. 

This linearity assumption is problematic in practice, as it omits two critical dynamics: diminishing marginal returns and the disproportionate impact that underperforming components may have on the system as a whole. To capture these dynamics, a more appropriate aggregation model is required; one that accounts for both the non-linear utility of improvements and the compounding effect of weaknesses.

Consider a set of four positive percentage scores,
\[
s_k \in [1,100], \qquad w_k > 0, \qquad k \in \{1, 2, 3, 4\}, \qquad \sum_{k=1}^{4} w_k = 1,
\]
where each $s_k$ denotes the score achieved in a specific CTI performance dimension, and $w_k$ represents its associated weight. A minimum floor value of 1 is imposed on the scores to prevent computational instability when evaluating logarithms. The four dimensions, quality, enrichment, integration, and operational impact, are respectively mapped to the vector
\[
\mathbf{s} = (s_1, s_2, s_3, s_4) = (Q_{\text{score}}, E_{\text{score}}, I_{\text{score}}, O_{\text{score}}).
\]

This paper introduces the \emph{Threat Intelligence Effectiveness Index} (TIEI), which is defined as the weighted geometric mean of these four components:
\begin{equation}
\label{eq:TIEI}
\mathrm{TIEI}
= 100 \prod_{k=1}^{4} \left( \frac{s_k}{100} \right)^{w_k}
= \prod_{k=1}^{4} s_k^{w_k},
\qquad \mathrm{TIEI} \in [1,100].
\end{equation}

This formulation offers several advantages. First, the geometric mean is more sensitive to low-performing dimensions than the arithmetic mean, thereby penalizing unbalanced performance and reducing compensability across dimensions. Second, it is normalized, which makes it well-suited for tracking CTI program performance over time within the same organization. Cross-organizational comparison is only meaningful when a common scoring standard (including shared weights, targets, and normalization rules) is applied.

Weighting each dimension $w_k$ requires a well-defined approach. A multi-stakeholder elicitation process is recommended, such as an Analytic Hierarchy Process (AHP) workshop or a budget allocation exercise, to derive these weights in a transparent and reproducible manner. The complete scoring specification, weights, targets, and normalization rules, should remain fixed throughout a longitudinal series. If an annual review changes that specification, prior scores should be recomputed from the retained raw data under the new rules. Where full restatement is not possible, the organization should calculate both versions for at least one overlap period, report the resulting bridge, and mark a structural break rather than treating the two series as directly comparable.

A known limitation of the geometric mean is its collapse to zero in the presence of zero-valued components: if any raw (unfloored) score were zero, the entire index would vanish regardless of the remaining dimensions. This motivates the floor of 1 already imposed on all scores in the definition of $\mathbf{s}$, which guarantees $\mathrm{TIEI} \in [1,100]$. In exceptional cases where a raw score of zero accurately reflects the complete absence of a capability, analysts should document this explicitly alongside the index rather than within the core calculation.

\paragraph{Aggregation baselines and weakest-link behaviour.}
Table~\ref{tab:tiei-designed-comparison} and Figure~\ref{fig:tiei-sensitivity} present the designed integration-stall comparison and its sensitivity across the full range of possible Integration scores.
To make the choice of geometric aggregation explicit rather than axiomatic, we compare TIEI with three baselines under the reference weights
$\mathbf{w}=(0.40,0.20,0.25,0.15)$:
\begin{align*}
A(\mathbf{s})&=\sum_k w_ks_k, &
G(\mathbf{s})&=\prod_k s_k^{w_k},\\
H(\mathbf{s})&=\left(\sum_k\frac{w_k}{s_k}\right)^{-1}, &
W(\mathbf{s})&=\min_k s_k.
\end{align*}

\begin{figure}[H] 
  \centering
  
  \captionsetup{type=table}
  \caption{Aggregation comparison for the designed integration-stall scenario.}
  \label{tab:tiei-designed-comparison}
  \begin{tabular}{lrrr}
  \toprule
  \textbf{Aggregator} & $\mathbf{s}^{(1)}$ & $\mathbf{s}^{(2)}$ & \textbf{Drop} \\
  \midrule
  Arithmetic          & 76.50 & 66.50 & $-13.07\%$ \\
  Geometric (TIEI)    & 75.59 & 57.44 & $-24.02\%$ \\
  Harmonic            & 74.65 & 46.02 & $-38.35\%$ \\
  Minimum             & 60.00 & 20.00 & $-66.67\%$ \\
  \bottomrule
  \end{tabular}

  \vspace{1.5em} 
  
  \captionsetup{type=figure} 
  \begin{tikzpicture}
    \begin{axis}[
      width=0.9\columnwidth,       
      height=0.65\columnwidth, 
      xmin=0, xmax=100,
      ymin=0, ymax=90,
      xtick={0,20,...,100},
      ytick={0,15,...,90},
      xlabel={Integration score (\%)},
      ylabel={Aggregate score},
      label style={font=\footnotesize},
      tick label style={font=\scriptsize},
      grid=major,
      grid style={dashed,gray!35},
      axis line style={black!70},
      legend style={
        at={(0.97,0.03)},
        anchor=south east,
        font=\tiny,
        draw=gray!45,
        fill=white,
        fill opacity=0.92,
        text opacity=1,
        cells={anchor=west}
      },
      legend cell align={left},
      samples=160,
      domain=1:100,
      no markers,
    ]
      \addplot[orange!90!black,thick]
        {61.5 + 0.25*x};
      \addlegendentry{Arithmetic}

      \addplot[red!80!orange,thick]
        {85^0.40 * 70^0.20 * x^0.25 * 90^0.15};
      \addlegendentry{Geometric (TIEI)}

      \addplot[blue!75!black,thick,dashed]
        {1/(0.40/85 + 0.20/70 + 0.25/x + 0.15/90)};
      \addlegendentry{Harmonic}

      \addplot[black!75,thick,densely dotted]
        {min(x,70)};
      \addlegendentry{Minimum}
    \end{axis}
  \end{tikzpicture}
  \caption{Sensitivity of four aggregation rules to \emph{Integration} score variations, with all other dimensions held constant. }
  \label{fig:tiei-sensitivity}
\end{figure}

Here $A$ permits full compensation, $H$ imposes a stronger low-score penalty,
and $W$ is a strict weakest-link rule. For a baseline profile
$\mathbf{s}^{(1)}=(85,70,60,90)$ and an otherwise identical profile with
stalled integration $\mathbf{s}^{(2)}=(85,70,20,90)$, the geometric mean occupies a transparent middle position between the fully offsetting and more severe alternatives.

\paragraph{Computational aggregation robustness.}
The designed example is complemented by a reproducible stress test of
$100{,}000$ four-dimensional profiles independently sampled from
$\mathrm{Uniform}(1,100)$ (seed 42).  For each alternative, we calculate Spearman rank correlation with TIEI, the share of independently drawn program pairs whose ordering reverses, and the median absolute score difference.
Table~\ref{tab:tiei-aggregator-robustness} shows high but non-equivalent rankings: the arithmetic and harmonic means reverse 10.74\% and 8.92\% of pairwise decisions, respectively, while the strict minimum reverses 18.34\%.

\begin{table}[ht]
\centering
\caption{Aggregator robustness over $100{,}000$ synthetic profiles.}
\label{tab:tiei-aggregator-robustness}
\begin{tabular}{lrrr}
\toprule
\textbf{Aggregator} & $\rho$ & \textbf{Reversal} & $\widetilde{|\Delta|}$ \\
\midrule
Arithmetic       & 0.932 & 10.74\% & 6.51 \\
Geometric (TIEI) & 1.000 & 0.00\%  & 0.00 \\
Harmonic         & 0.950 & 8.92\%  & 5.86 \\
Minimum          & 0.826 & 18.34\% & 20.49 \\
\bottomrule
\end{tabular}
\end{table}

\paragraph{Sensitivity to top-level weights and score floor.}
We next compare the reference weights with equal weights, an
Operational-Impact-heavy specification $(0.20,0.15,0.20,0.45)$, and an
Integration-heavy specification $(0.20,0.15,0.45,0.20)$. Maturity classes are
defined by cut points at 25, 50, and 75. As shown in
Table~\ref{tab:tiei-weight-robustness}, rankings remain positively associated,
but classification and pairwise decisions can change materially under strongly
different priorities. The weights must therefore be elicited transparently and
held fixed for longitudinal comparisons.

\begin{table}[ht]
\centering
\caption{TIEI sensitivity to alternative top-level weights.}
\label{tab:tiei-weight-robustness}
\begin{tabular}{lrrr}
\toprule
\textbf{Weights} & $\rho$ & \textbf{Reversal} & \textbf{Class change} \\
\midrule
Equal             & 0.944 & 10.33\% & 17.72\% \\
Operational-heavy & 0.804 & 19.49\% & 34.53\% \\
Integration-heavy & 0.880 & 15.20\% & 27.14\% \\
\bottomrule
\end{tabular}
\end{table}

As a continuous uncertainty check, weights drawn from
$\mathrm{Dirichlet}(20\mathbf{w})$ retain the reference vector as their mean.
They produce a median absolute score difference of 3.01 points, a 95th-percentile
difference of 11.41 points, and 8.89\% pairwise reversals relative to the fixed
reference weights.

Finally, we repeat the stress test with raw profiles sampled over $[0,100]$ and
floors of 0.1, 1, and 5. Relative to the default floor of 1, lowering the floor
to 0.1 changes 3.97\% of scores but only 0.35\% of maturity classes; raising it
to 5 changes 18.53\% of scores, yields a 95th-percentile absolute difference of
6.24 points, and changes 2.70\% of classes. Thus the floor has limited aggregate
effect in the full design space but can matter for near-zero capabilities and must be disclosed \footnote{The complete calculation scripts are provided at https://zenodo.org/records/
21319607}.

\paragraph{Sub-Metric Scoring Schema and Normalization Procedures for the TIEI}
\label{sec:tiei-scoring-rubric}

To ensure transparent and reproducible application of the Threat Intelligence Effectiveness Index (TIEI), this subsection formalizes the scoring framework for its four components, as previously introduced in equation~\ref{eq:TIEI}. The schema uses measures obtainable from ordinary CTI, SIEM/SOAR, CMDB, ticketing, and incident-response records. Before each measurement period, the organization must register the observation window, targets, data sources, evidence rules, and the populations designated as eligible, priority, or critical. These definitions remain fixed during the period so that changes in scope cannot artificially improve a score.

Each sub-metric is normalized to a $[1,100]$ scale relative to an internally defined performance target derived from historical baselines, peer benchmarks, or service-level commitments. For a positive attainment rate $x$, the default normalization is $B(100x/t)$; for a lower-is-better duration $h$, it is $B(100t/h)$, where $t$ is the corresponding target and $B(z)=\max(1,\min(100,z))$. When no defensible universal target is available, Appendix~\ref{app:multi-anchor-normalization} provides an optional multi-anchor normalization that preserves the TIEI scale and aggregation rule. For a delta-based improvement $d$, let $r<0$ denote the predefined regression threshold and $t>0$ the target improvement. The normalization is
\[
N_{\Delta}(d;r,t)=B\!\left(\begin{cases}
1+49(d-r)/(-r), & r\le d\le 0,\\
50+50d/t, & 0<d\le t.
\end{cases}\right),
\]
with values below $r$ assigned 1 and values above $t$ assigned 100. Thus regression, stagnation, and target attainment remain distinguishable. Every reported score must include its numerator, denominator, observation window, and sample size; matched comparisons must use the same threat category and severity band.

For the duration-based operational metrics, positive values represent reductions relative to a strictly positive baseline:
\begin{align*}
d_{\mathrm{MTTD}}
&=100\frac{\mathrm{MTTD}_0-\mathrm{MTTD}_1}{\mathrm{MTTD}_0},\\
d_{\mathrm{MTTR}}
&=100\frac{\mathrm{MTTR}_0-\mathrm{MTTR}_1}{\mathrm{MTTR}_0},
\end{align*}
where subscripts 0 and 1 denote the registered baseline and comparison periods. If a baseline is zero, the relative change is undefined and must not be scored. Open or right-censored incidents must be handled using a prespecified censoring-aware method, such as survival analysis, and the method and number of censored observations must be reported.

Tables~\ref{tab:tiei-Q}--\ref{tab:tiei-O} present the sub-metrics associated with each TIEI component, together with their raw performance measures, normalization methods, and weights within the corresponding component score.

\newcounter{tieisubmetricbase} 
\setcounter{tieisubmetricbase}{\value{table}}
\begin{table*}[t]
  \centering
  \footnotesize
  \begin{minipage}[t]{0.485\textwidth}
  \vspace{0pt}
  \captionsetup{type=table,width=\linewidth,justification=raggedright}
  \caption{Intelligence Quality (Q) Sub-metrics.}
  \label{tab:tiei-Q}
  \setlength{\extrarowheight}{0.48ex}
  \begin{tabularx}{\linewidth}{@{}l >{\raggedright\arraybackslash}X c@{}}
    \toprule
    Sub-metric & Raw Measure\textsuperscript{*} & Weight \\
    \midrule
    Accuracy & $\%$ of sampled items validated as correct (not false or retracted) in periodic review; normalized as $(\text{rate}/\text{target}_{\mathrm{acc}})\times100$ & 0.35 \\
    Timeliness & Median hours from external observation to internal availability; normalized as $(\text{target}_{\mathrm{time}}/\text{hrs})\times100$ & 0.25 \\
    Relevance & $\%$ of items linked to an asset, vulnerability, business process, or priority actor/TTP; normalized as $(\text{rate}/\text{target}_{\mathrm{rel}})\times100$ & 0.25 \\
    Completeness & $\%$ of items with all mandatory fields populated under the registered schema; normalized as $(\text{rate}/\text{target}_{\mathrm{comp}})\times100$ & 0.15 \\
    \bottomrule
  \end{tabularx}
  \end{minipage}%
  \hfill%
  \begin{minipage}[t]{0.485\textwidth}
  \vspace{0pt}
  \setcounter{table}{\value{tieisubmetricbase}}
  \addtocounter{table}{2}
  \captionsetup{type=table,width=\linewidth,justification=raggedright}
  \caption{Integration \& Automation (I) Sub-metrics.}
  \label{tab:tiei-I}
  \setlength{\extrarowheight}{0pt}
  \begin{tabularx}{\linewidth}{@{}l >{\raggedright\arraybackslash}X c@{}}
    \toprule
    Sub-metric & Raw Measure & Weight \\
    \midrule
    Control Coverage & $\%$ of critical security tools or workflows consuming CTI; normalized as $(\text{rate}/\text{target}_{\mathrm{cov}})\times100$ & 0.25 \\
    Delivery Success Rate & $\%$ of CTI pipeline jobs completed without ingestion, parsing, or delivery failure; normalized as $(\text{rate}/\text{target}_{\mathrm{del}})\times100$ & 0.25 \\
    Automation Rate & $\%$ of eligible items automatically actioned in security workflows; normalized as $(\text{rate}/\text{target}_{\mathrm{auto}})\times100$ & 0.30 \\
    SLA Adherence & $\%$ of enrichment or dissemination events completed within SLA; normalized as $(\text{rate}/\text{target}_{\mathrm{sla}})\times100$ & 0.20 \\
    \bottomrule
  \end{tabularx}
  \end{minipage}%

  \vspace{3.25ex}

  \begin{minipage}[t]{0.485\textwidth}
  \vspace{0pt}
  \setcounter{table}{\value{tieisubmetricbase}}
  \addtocounter{table}{1}
  \captionsetup{type=table,width=\linewidth,justification=raggedright}
  \caption{Enrichment (E) Sub-metrics.}
  \label{tab:tiei-E}
  \setlength{\extrarowheight}{0pt}
  \begin{tabularx}{\linewidth}{@{}l >{\raggedright\arraybackslash}X c@{}}
    \toprule
    Sub-metric & Raw Measure & Weight \\
    \midrule
    TTP Mapping Rate & $\%$ of behavior-containing items mapped to ATT\&CK techniques or sub-techniques; normalized as $(\text{rate}/\text{target}_{\mathrm{ttp}})\times100$ & 0.25 \\
    Internal Correlation Rate & $\%$ of items correlated to internal telemetry, assets, cases, or vulnerabilities; normalized as $(\text{rate}/\text{target}_{\mathrm{corr}})\times100$ & 0.35 \\
    Entity-Linking Rate & $\%$ of items with recorded actor, campaign, malware, CVE, tool, or infrastructure links; normalized as $(\text{rate}/\text{target}_{\mathrm{ent}})\times100$ & 0.20 \\
    Actionability Rate & $\%$ of items linked to a detection, playbook, ticket, blocklist, or advisory; normalized as $(\text{rate}/\text{target}_{\mathrm{act}})\times100$ & 0.20 \\
    \bottomrule
  \end{tabularx}
  \end{minipage}%
  \hfill%
  \begin{minipage}[t]{0.485\textwidth}
  \vspace{0pt}
  \setcounter{table}{\value{tieisubmetricbase}}
  \addtocounter{table}{3}
  \captionsetup{type=table,width=\linewidth,justification=raggedright}
  \caption{Operational Impact (O) Sub-metrics.}
  \label{tab:tiei-O}
  \setlength{\extrarowheight}{0pt}
  \begin{tabularx}{\linewidth}{@{}l >{\raggedright\arraybackslash}X c@{}}
    \toprule
    Sub-metric & Raw Measure & Weight \\
    \midrule
    Detection Lift & $d_{\mathrm{MTTD}}$ for CTI-enabled detections versus baseline; normalized with $N_{\Delta}$ & 0.30 \\
    \addlinespace[2.5625ex]
    Response Lift & $d_{\mathrm{MTTR}}$ for CTI-enabled cases versus baseline; normalized with $N_{\Delta}$ & 0.25 \\
    \addlinespace[2.5625ex]
    False-Positive Reduction & $\%$ reduction in false positives within CTI-enriched alert streams; normalized with $N_{\Delta}$ & 0.20 \\
    \addlinespace[2.5625ex]
    Coverage Lift & $\%$ increase in validated coverage of priority TTPs or threat scenarios; normalized with $N_{\Delta}$ & 0.25 \\
    \addlinespace[3.3125ex]
    \bottomrule
  \end{tabularx}
  \end{minipage}
  \setcounter{table}{\value{tieisubmetricbase}}
  \addtocounter{table}{4}

  \vspace{1ex}
  \parbox{\textwidth}{\raggedright\scriptsize * All normalized scores ($x$) are bounded to $[1,100]$ using $B(x)=\max(1,\min(100,x))$ before geometric aggregation.}
\end{table*}

Each set of sub-metrics is aggregated using a weighted geometric mean, with weights summing to 1. Because empirical data dictating the exact universal importance of each sub-metric does not currently exist, we establish a reference baseline based on domain-expert heuristics and proximity-to-impact. Specifically, higher weights are given to factors that most directly affect downstream security decisions and operational outcomes, while lower weights are assigned to supporting controls that improve consistency, completeness, or process reliability. To maintain mathematical consistency with the top-level TIEI formulation (Equation~\ref{eq:TIEI}), a default floor value of 1 is applied to all normalized sub-metric scores to prevent a zero-value collapse. The resulting component scores $S_Q, S_E, S_I, S_O$ are expressed on a standardized 1-100 scale. The overall TIEI is then computed as the weighted geometric mean as shown in Equation~\ref{eq:TIEI}.
Notably, the sub-metrics across these dimensions are not strictly statistically independent; for instance, high Completeness (Table~\ref{tab:tiei-Q}) supports a high Automation Rate (Table~\ref{tab:tiei-I}). However, the TIEI is designed as a pipeline maturity index rather than a predictive regression model, meaning this covariance is an intended reflection of the CTI lifecycle rather than a statistical flaw. While these metrics correlate, they measure distinct operational phases that can diverge.

\subsection{Qualitative Value Indicators}
\label{qualitative}
\-
\newline
Qualitative indicators, while less easily monetized, often capture the most strategic benefits of CTI; they address executive decision‑making, brand resilience, and regulatory posture.

\paragraph{Strategic Decision‑Making and Risk Management} CTI provides leadership with a contextualized understanding of the threat landscape, supporting decisions on capital allocation, vendor selection, and risk tolerance. A key application is threat‑informed vulnerability management. By overlaying CTI with vulnerability data, security teams prioritize remediation based on real-world exploitation rather than theoretical severity, shifting patch management from just compliance‑driven to risk‑driven \cite{SANSCTIDefinition}.

\paragraph{Brand and Reputation Protection} Brand equity is an intangible asset that can erode rapidly after publicized breaches or brand‑spoofing campaigns. Through continuous monitoring of the Internet, dark web, and mobile app stores for malicious look‑alikes, fraudulent domains, and counterfeit products, it is possible to detect these events \cite{RecordedFutureBrandReputation}. Key performance indicators (KPIs) include the volume and dwell time of takedown requests, reductions in fraudulent customer contacts, and changes in third‑party brand‑health scores. Although direct monetary valuation varies by industry and over time, these proxy metrics provide longitudinal indicators of reputational value.

\paragraph{Regulatory Compliance and Contractual Advantage} Demonstrating proactive cyber‑risk management can unlock contractual opportunities and reduce regulatory examination. CTI feeds the risk assessments mandated by frameworks and regulations such as the NIST Cybersecurity Framework (CSF), the EU's Digital Operational Resilience Act (DORA), and sectoral regulations like HIPAA \cite{NISTCSF2024, DORA2022, ENISAFinance2025}. Evidence of a mature CTI capability often features as a differentiator in competitive tenders, with measurable influence on contract awards and insurance premiums. Although the financial impact of such advantages will vary, security leaders can document the qualitative uplift using procurement feedback, auditor reports, and insurer questionnaires.

A summary of all the qualitative and quantitative measurement methodologies and the indicators discussed is provided in Appendix \ref{app:roi-indicator-tables}.

\paragraph{Quantifying Detection Coverage and Mitigation Effectiveness}

The MITRE ATT\&CK framework provides a globally recognized taxonomy of adversary TTPs. CTI operationalizes ATT\&CK by mapping intelligence reports to specific techniques, enabling security teams to prioritize defenses against behaviors most relevant to their threat profile \cite{mitre_attack_website, infosec_attack_cti}.

Operationalization begins with threat-informed prioritization: CTI identifies the ATT\&CK techniques most frequently observed in campaigns targeting the organization's industry and region. Existing controls are then mapped to those priority techniques in ATT\&CK Navigator, allowing detection and prevention gaps to be identified without treating all techniques as equally relevant. The resulting gaps guide the implementation of analytics, security-control changes, and response playbooks, whose effectiveness should be validated through adversary emulation or purple-team exercises. Finally, the organization reports changes in coverage against the same prioritized technique set and links validated improvements to risk-reduction estimates generated through FAIR or a comparable model. Keeping the threat scope, control mapping, and validation method stable across measurement periods makes the resulting coverage trend interpretable. By combining operational evidence (ATT\&CK), governance context (NIST CSF), and financial translation (FAIR), CTI teams create a layered ROI narrative that preserves its relevance and credibility as sector-specific data evolves.

TIEI should be assessed at defined intervals to incorporate new threat intelligence, changing business priorities, and evolving regulatory requirements. A trendline is interpretable only while the scoring specification remains fixed or historical scores have been restated or bridged as described above.

%% file: Sections/03-UseCases.tex
\section{Sector-Specific Feasibility Analysis}
\label{ch:sector_rois}

This section applies the financial framework to two sector-specific scenarios: financial services and healthcare. The analysis proceeds in three stages. First, it derives a common breakeven frontier in the two organization-specific unknowns: the annual probability of a scope-matched material event and the mitigation attributable to the proposed incremental CTI capability. Second, it shows how uncertainty in mitigation can be propagated conditionally once a decision-maker supplies a defensible event probability. Third, it applies the same sequence to each sector: establish the threat and CTI context, document the loss and cost inputs, compute the feasibility frontier, and interpret the result alongside complementary value indicators. This structure separates empirical inputs, modeling assumptions, and decision implications while enabling a like-for-like comparison of the two scenarios.

The analysis evaluates the economic conditions under which CTI can create value; it does not assign a sector-wide probability to a material loss event. Public reports generally measure broader populations and outcomes, for example, whether an organization experienced any cyberattack or ransomware activity, and do not identify the counterfactual probability of the narrower material event represented by a breach-cost benchmark \cite{Brightmore2025,Sophos2024,IBM2025}. Treating those prevalence figures as "without CTI" loss probabilities would therefore create event-scope and causal mismatches.

\subsection{Breakeven Feasibility Frontier}

For one clearly defined material event class, let $p_0\in[0,1]$ be its annual probability under existing controls but before the proposed incremental CTI capability, $LM$ its conditional loss magnitude, and $M\in[0,1]$ the fraction of baseline Annualized Loss Expectancy (ALE) reduced by CTI-informed controls. Both $p_0$ and $M$ are organization-specific unknowns. The first must be assessed from scope-matched internal evidence or elicitation; the second must be attributed to the incremental CTI capability. All costs are evaluated over the same one-year decision horizon. Then
\begin{equation}\label{eq:ale_baseline}
\mathrm{ALE}_0=p_0LM,
\qquad
\Delta\mathrm{ALE}=p_0MLM.
\end{equation}
Using the all-in annual incremental investment $\mathrm{TCO}_{\text{CTI}}$ defined in Equation~\eqref{eq:CTICost}, ROI is
\begin{equation}\label{eq:roi_cti}
R_{\text{CTI}}=\frac{p_0MLM-\mathrm{TCO}_{\text{CTI}}}{\mathrm{TCO}_{\text{CTI}}}.
\end{equation}

Define the observable scenario ratio
\begin{equation}\label{eq:kappa}
\kappa=\frac{\mathrm{TCO}_{\text{CTI}}}{LM}.
\end{equation}
Positive ROI is equivalent to
\begin{equation}\label{eq:breakeven_frontier}
R_{\mathrm{CTI}}>0
\quad\Longleftrightarrow\quad
p_0M>\kappa.
\end{equation}
Equation~\eqref{eq:breakeven_frontier} defines a feasibility frontier in the two unknowns $M$ and $p_0$. Any pair satisfying $M>\kappa/p_0$ yields a positive avoided-loss ROI, whereas pairs below this frontier do not. The corresponding threshold values are $p_0^{*}=\kappa/M$ for a given internally supported value of $M$, and $M^{*}=\kappa/p_0$ for a given internally supported value of $p_0$. Since $p_0\leq1$ and $M\leq1$, positive ROI is infeasible in this single-event avoided-loss model when $\kappa\geq1$. For repeatable event classes, the same derivation applies with the annual event frequency $\lambda_0$ replacing $p_0$.

\begin{table}[H]
\centering
\caption{Sector feasibility constants. $LM$ and $\mathrm{TCO}_{\mathrm{CTI}}$ are in millions of dollars and millions of dollars per year, respectively. $\mathrm{TCO}_{\mathrm{CTI}}$ is the all-in annual incremental investment defined in Equation~\eqref{eq:CTICost}. The endpoint range varies that cost by $\pm20\%$ and uses the two reported IBM loss magnitudes.}
\label{tab:breakeven-first}
\begin{tabularx}{\columnwidth}{@{}l*{4}{Y}@{}}
\toprule
\textbf{Sector} & \textbf{$LM$} & \textbf{TCO} & \textbf{$\kappa$} & \textbf{$\kappa$ range} \\
\midrule
Finance    & 5.820 & 0.40 & 6.87\% & 5.26--8.63\% \\
Healthcare & 8.595 & 0.60 & 6.98\% & 4.91--9.70\% \\
\bottomrule
\end{tabularx}
\end{table}

The numerical costs in Table~\ref{tab:breakeven-first} are therefore $\mathrm{TCO}_{\mathrm{CTI}}$ scenario envelopes, not platform prices. A reported or budgeted amount that covers only the CTI team, feeds, or platform must not be substituted directly for $\mathrm{TCO}_{\mathrm{CTI}}$. The adopting organization must instead calculate the complete investment using Equation~\eqref{eq:CTICost}, including the downstream enablement included in $C_{\text{integration}}$. Omitting any of these components moves the frontier downward and overstates feasibility.

Because the finance and healthcare values of $\kappa$ differ by only 0.11 percentage points, Table~\ref{tab:frontier_pairs} reports the boundaries numerically, translating Equation~\eqref{eq:kappa} into example pairs. These values are not estimates of sector risk or CTI effectiveness; they show what one unknown would have to equal at selected values of the other.

\begin{table}[H]
\centering
\caption{Minimum annual material-event probability $p_0^{*}=\kappa/M$ at selected CTI-attributable mitigation values.}
\label{tab:frontier_pairs}
\begin{tabularx}{\columnwidth}{@{}l*{5}{Y}@{}}
\toprule
\textbf{Sector} & \mbox{\textbf{$M=10\%$}} & \mbox{\textbf{$M=20\%$}} & \mbox{\textbf{$M=30\%$}} & \mbox{\textbf{$M=40\%$}} & \mbox{\textbf{$M=50\%$}} \\
\midrule
Finance    & 68.7\% & 34.4\% & 22.9\% & 17.2\% & 13.7\% \\
Healthcare & 69.8\% & 34.9\% & 23.3\% & 17.5\% & 14.0\% \\
\bottomrule
\end{tabularx}
\end{table}

For example, the finance scenario at $M=20\%$ breaks even only if the organization's scope-matched annual material event probability exceeds 34.4\%.

\subsection{Conditional Mitigation Analysis}
\label{sec:mit_factors}
The parameter $M$ is inherently organization-specific: its magnitude depends on CTI program maturity, the depth of control integration, and the prevailing sectoral threat environment. Consequently, no single cross-organizational value of $M$ can be treated as universally valid, and no independent study currently provides a general empirical estimate of the fractional reduction in Annualized Loss Expectancy attributable specifically to CTI-informed controls. Existing reports instead quantify related indicators. IBM associates threat intelligence with an average breach-cost reduction of \$211{,}906 \cite{IBM2025}, while the SANS 2025 CTI Survey reports that approximately 70\% of organizations observed improved detection and response after CTI integration \cite{SANS2025Survey}. Mandiant reports EMEA median dwell times of 32~days for externally notified events and 20~days for internally discovered events, a 12-day difference \cite{MandiantDwellTime2025}. Separately, IBM reports a 61-day shorter data-breach lifecycle for internally detected breaches relative to attacker-disclosed breaches \cite{IBM2024}. The Mandiant statistic concerns dwell time, whereas IBM's concerns the broader breach lifecycle. These associations motivate modeling risk reduction but do not identify $M$.

Given this evidence structure, adopting a single deterministic estimate for $M$ would introduce spurious precision. For the secondary illustration, we represent $M$ as a single net factor rather than estimating separate CTI effects on LEF and LM, and model it using three elicited scenario points (pessimistic, most likely, optimistic) within a PERT distribution, consistent with FAIR's probabilistic framework \cite{FAIRInstituteWhatIsFAIR,ConradMonteCarlo}. This simplification does not model channel-specific effects or their correlation. The PERT values are calibration assumptions for a reference organization at a specified maturity level, rather than universal constants or evidence that CTI will clear the breakeven threshold. Specifically, the bounds are conditioned on the program's TIEI, as formalized in Section~\ref{sec:tiei-bridge}, and are defined as zero-floored affine functions of a normalized TIEI score:

\begin{align}
    l(\mathrm{TIEI})
        &= \max\!\left\{0,\; l_0 \;+\; \alpha_l \;\cdot\;
           \frac{\mathrm{TIEI} - 1}{99}\right\},
        \label{eq:pert-l} \\[4pt]
    m(\mathrm{TIEI})
        &= \max\!\left\{0,\; m_0 \;+\; \alpha_m \;\cdot\;
           \frac{\mathrm{TIEI} - 1}{99}\right\},
        \label{eq:pert-m} \\[4pt]
    u(\mathrm{TIEI})
        &= \max\!\left\{0,\; u_0 \;+\; \alpha_u \;\cdot\;
           \frac{\mathrm{TIEI} - 1}{99}\right\},
        \label{eq:pert-u}
\end{align}
where $\tfrac{\mathrm{TIEI}-1}{99} \in [0,1]$ normalizes the index to the unit interval, and the $\max\{0,\cdot\}$ operator enforces the non-negativity floor directly in the model. The base parameters $(l_0, m_0, u_0)$ are the unconstrained affine intercepts before flooring, and the slope coefficients $(\alpha_l, \alpha_m, \alpha_u)$ control how rapidly each bound increases as program maturity improves. This formulation creates a direct, formal link between the operational performance metric introduced in Section~\ref{sec:tiei-bridge} and the financial ROI model: a program's measured quality (TIEI) determines the range of plausible risk-reduction outcomes ($M_i$), while the residual spread of the PERT distribution captures factors external to the TIEI measurement, such as variability in attacker sophistication and organizational culture.

For simulation, we use the modified PERT parameterization with shape factor $\lambda=4$ \cite{ConradMonteCarlo, vose2008risk}:
\begin{align}
    a_{\mathrm{PERT}} &= 1 + \lambda\frac{m-l}{u-l},
    & b_{\mathrm{PERT}} &= 1 + \lambda\frac{u-m}{u-l},
    \label{eq:pert-shape}\\[2pt]
    M_i &= l + (u-l)X,
    & X &\sim \mathrm{Beta}(a_{\mathrm{PERT}}, b_{\mathrm{PERT}}),
    \label{eq:pert-beta}
\end{align}
with $l=l(\mathrm{TIEI})$, $m=m(\mathrm{TIEI})$, and $u=u(\mathrm{TIEI})$.

We calibrate the six parameters using two illustrative reference scenarios: a lower-maturity program ($\mathrm{TIEI} \approx 25$) with modest integration, yielding $\mathrm{PERT} \approx (0.03,\; 0.15,\; 0.28)$, and a higher-maturity program ($\mathrm{TIEI} \approx 80$) with full SOAR automation and proactive threat hunting, yielding $\mathrm{PERT} \approx (0.10,\; 0.40,\; 0.60)$. These triplets function as elicited calibration anchors for the simulation model and are not universally applicable mitigation rates. Solving the resulting linear system for the unconstrained affine terms, then applying Equations~\eqref{eq:pert-l}--\eqref{eq:pert-u}, gives the default parameter set shown in Table~\ref{tab:pert-calibration}.

\begin{table}[ht]
\centering
\caption{Default calibration of TIEI-conditioned PERT parameters. Values are obtained by solving the linear system defined by the two anchor points ($\mathrm{TIEI}=25$ and $\mathrm{TIEI}=80$). Reported coefficients are rounded for presentation; the effective bounds are computed via Equations~\eqref{eq:pert-l}--\eqref{eq:pert-u}.}
\label{tab:pert-calibration}
\begin{tabular}{l c c c}
\toprule
\textbf{Parameter} & \textbf{Base ($\cdot_0$)} & \textbf{Slope ($\alpha$)} & \textbf{Interpretation} \\
\midrule
$l$ (pessimistic)   & $-0.0005$ & 0.1260 & Lower bound  \\
$m$ (most likely)    & 0.0409 & 0.4500 & Central estimate \\
$u$ (optimistic)     & 0.1404 & 0.5760 & Upper bound \\
\bottomrule
\end{tabular}
\end{table}

\noindent For a conditional illustration, we assume a mature CTI capability corresponding to $\mathrm{TIEI} = 80$, which under this calibration approximately recovers (and equals when unrounded coefficients are used):
\[
  M_i \;\sim\; \mathrm{PERT}\!\left(l(80),\; m(80),\; u(80)\right)
      \;\approx\; \mathrm{PERT}(0.10,\; 0.40,\; 0.60).
\]
The three points ($l = 0.10$, $m = 0.40$, $u = 0.60$) are modeling assumptions selected for simulation purposes. Because CTI mitigation effectiveness is strictly organization-dependent (e.g., program maturity, integration depth, operating model, and sector threat exposure), assigning a single deterministic value would imply false precision. We therefore represent $M_i$ with a pessimistic lower bound, a most-likely central value, and an optimistic upper bound to span a scenario range for a mature program. Holding TIEI fixed at 80 controls the mitigation calibration across sectors; $LM$ and $\mathrm{TCO}_{\text{CTI}}$ vary by sector, while $p_0$ remains an explicitly selected stress-test input. Organizations with different TIEI assessments can substitute their own score into Equations~\eqref{eq:pert-l}--\eqref{eq:pert-u}; for instance, a program at $\mathrm{TIEI} = 50$ would yield $\mathrm{PERT}(0.06,\; 0.26,\; 0.43)$, while one at $\mathrm{TIEI} = 95$ would yield $\mathrm{PERT}(0.12,\; 0.47,\; 0.69)$.

A Monte Carlo simulation of $N = 100{,}000$ iterations draws only $M_i$ from the PERT distribution\footnote{The simulation code is available at \url{https://zenodo.org/records/21319607}; the feasibility calculations and machine-readable results are included with the accompanying reproducibility materials.}. It does not draw or estimate $p_0$. Instead, Table~\ref{tab:conditional-positive-roi} conditions on three transparent stress-test values of $p_0$ and reports
\begin{equation}\label{eq:conditional_success}
\Pr(R_{\mathrm{CTI}}>0\mid p_0)
=\Pr\!\left(M_i>\frac{\kappa}{p_0}\right).
\end{equation}

For $M_i\sim\mathrm{PERT}(0.10,0.40,0.60)$, the sampled distribution has median 38.7\% and a 90\% simulation interval of approximately [22.2\%, 53.0\%].

\begin{table}[H]
\centering
\caption{Conditional probability of positive avoided-loss ROI at selected, non-estimated values of $p_0$ ($M_i\sim\mathrm{PERT}(0.10,0.40,0.60)$; $N=100{,}000$; seed = 42).}
\label{tab:conditional-positive-roi}
\begin{tabular}{lrrr}
\toprule
\textbf{Sector} & \textbf{$p_0=20\%$} & \textbf{$p_0=40\%$} & \textbf{$p_0=60\%$} \\
\midrule
Finance    & 66.1\% & 99.0\% & 100.0\% \\
Healthcare & 64.3\% & 98.9\% & 100.0\% \\
\bottomrule
\end{tabular}
\end{table}

The two applications that follow use a common reporting structure so that their evidentiary basis and decision thresholds can be compared directly. They document the loss and cost inputs and interpret the resulting feasibility frontiers.

\subsection{Financial Services}\label{sec:finance_roi}
\paragraph{Threat context and CTI pathways.}
The financial services sector is a major adopter of CTI, driven by high stakes and strict regulatory expectations \cite{fsisac2025navigating}. Banks, investment firms, and insurance companies face advanced persistent threats (APTs) (e.g.,\ nation-state groups targeting payment networks) and organized cybercrime (e.g., groups exploiting online banking and ATM systems). In this environment, CTI programs typically ingest threat data from industry sharing groups (e.g.,\ FS-ISAC), commercial intelligence feeds, and law-enforcement alerts, converting them into actionable insights for security operations. Key applications include fraud-indicator tracking, attribution of phishing campaigns aimed at customers, and strategic intelligence on geopolitical threats to the financial system. Mapping CTI outputs to MITRE ATT\&CK enables financial organizations to assess defensive coverage against the Tactics, Techniques, and Procedures (TTPs) most relevant to their threat landscape \cite{infosec_attack_cti}. For example, threat intelligence on new SWIFT payment-fraud malware can be mapped to corresponding ATT\&CK techniques (e.g.,\ \textit{Valid Accounts} for stolen credentials and \textit{Ingress Tool Transfer} for malware delivery) and used to harden controls \cite{infosec_attack_cti}.

Reported evidence associates CTI integration with improvements in incident detection and response. For instance, a SANS survey found that $\sim$70\% of organizations reported enhanced detection and response capabilities following CTI integration \cite{SANS2025Survey}. Early-warning intelligence and indicators of compromise (IOCs) can support earlier identification of malicious activity, proactive hunting for TTP patterns, and more efficient incident response. Mandiant's \emph{M-Trends 2025} report, for example, reports a median dwell time of five days when ransomware becomes observable through encryption or a ransom note \cite{MandiantDwellTime2025}. This short interval illustrates the limited window available for proactive detection and mitigation.

\paragraph{Scenario inputs.}
An industry report states that 48\% of surveyed finance and insurance organizations experienced a cyberattack in the preceding 12 months \cite{Brightmore2025}. That statistic provides threat context but is not used as $p_0$: it neither isolates the material breach scope represented by IBM's loss estimate nor identifies risk in the absence of the proposed CTI increment. IBM reports average financial-industry breach costs of \$6.08~M in 2024 \cite{IBM2024}, also stated in IBM's finance summary \cite{Bonderud2024}, and \$5.56~M in 2025 \cite{IBM2025}; their equal-weight mean is \$5.82~M. The scenario inputs are therefore
\[
\begin{aligned}
LM &= (\$6.08\text{M}+\$5.56\text{M})/2 = \$5.82\;\text{million}, \\
\mathrm{TCO}_{\text{CTI}} &= \$0.40\;\text{million/year} \;\text{\cite{ENISA2022Fin}}.
\end{aligned}
\]

\paragraph{Feasibility result.}
The resulting feasibility constant and frontier are
\[
\kappa_f=\frac{0.40}{5.82}=0.0687,
\qquad
p_{0,f}M>0.0687.
\]
For example, $M=20\%$ requires $p_{0,f}>34.4\%$, while $p_{0,f}=20\%$ requires $M>34.4\%$. Neither value is asserted by the sector data; the frontier identifies what the organization must substantiate.

\paragraph{Complementary value indicators.}
A vendor-commissioned TEI study in the Recorded Future context reported a 245\% ROI over three years, alongside faster threat identification and reduced investigation effort \cite{IDC2021}. Beyond avoided losses, CTI can support executives in prioritizing security investments and adjusting controls to the threat landscape. For instance, threat-trend reports showing an increase in ransomware targeting core banking systems can motivate investment in specific mitigations (e.g., network isolation and backup drills), aligning security expenditure with observed risk. CTI can also inform scenario planning: intelligence on an APT group's tactics enables realistic red-team exercises that test resilience and guide strategic improvements.

CTI may also contribute to brand protection and customer trust, both of which are particularly salient for financial institutions. Relevant proxy metrics, discussed in Section~\ref{qualitative}, include cyber-insurance premiums and customer retention. A strong intelligence-led defense may contribute to lower insurance costs where insurers recognize demonstrable risk reduction and may reduce exposure to the public-relations and market consequences of a breach. Prior studies associate public breach disclosure with adverse stock-price reactions and increased customer attrition \cite{johnson2017stock, durongkadej2023data}. These outcomes are not included in the avoided-loss frontier, but they provide complementary indicators of business value.

\subsection{Healthcare}
\label{sec:healthcare_roi}
\paragraph{Threat context and CTI pathways.}
Cyber threats remain a consistent pressure on the healthcare sector, with ransomware identified as the single most significant threat, accounting for 54\% of all reported incidents. The U.S. Department of Health and Human Services (HHS) documented a 264\% increase in ransomware attacks against the sector between 2019 and 2024 \cite{healthisac_cti_box_2024}. The consequences extend beyond financial costs to patient safety: a ransomware attack on a Barcelona hospital, for example, forced the cancellation of 150 non-urgent operations and 3,000 patient checkups \cite{Barcelona_Hospital_Incident}. Cost avoidance is therefore a central component of the healthcare CTI scenario.

The sector's vulnerability is influenced by several key factors. First, a high dependency on external partners for services and technology creates a significant risk from third-party breaches. The ransomware attack on Change Healthcare in early 2024 serves as a clear example, halting pharmacy and billing operations nationwide and costing its parent company, UnitedHealth Group, over \$872 million in the first quarter alone \cite{Gatlan2024}. Second, the increasing use of the Internet of Medical Things (IoMT) has introduced a vast attack surface that is increasingly difficult to secure. Connected medical devices, which often have long operational lifespans and complex patching requirements, present unique vulnerabilities that attackers are likely to exploit \cite{Alder2025}. Finally, the human element remains a critical weakness, with phishing representing a primary initial access vector for major attacks \cite{Alder2024}.

\paragraph{Scenario inputs.}
Healthcare remains the costliest industry for breaches: IBM reports averages of \$9.77~M in 2024 \cite{IBM2024} and \$7.42~M in 2025 \cite{IBM2025}. These totals include not only IT recovery and fines (under regulations like HIPAA or GDPR) but also downstream costs such as patient notification, potential lawsuits, and loss of trust. A single successful ransomware attack can disrupt hospital operations; for example, the 2017 \textit{WannaCry} attack forced the UK National Health Service (NHS) to cancel 19,000 appointments and cost an estimated £92 million in direct and indirect losses \cite{CyberSecurityPolicyUK2018}. This example is used to illustrate impact magnitude rather than to claim a precise counterfactual prevention rate: timely vulnerability intelligence and patch coordination are plausible mitigation channels, but the avoidable fraction cannot be inferred directly from that historical incident.

Sophos reports that 67\% of surveyed healthcare organizations were affected by ransomware in its study period \cite{Sophos2024}. This broad victimization statistic is not used as $p_0$: "affected by ransomware" is not equivalent to the annual probability of an IBM-scope material breach under existing controls but before an incremental CTI capability. As a transparent loss-magnitude proxy, we pool IBM's 2024 and 2025 global healthcare averages, giving \$8.595~M per breach \cite{IBM2025}. The scenario assumes a \$0.60~M all-in annual investment, including downstream enablement. 
Therefore,
\[
\begin{aligned}
LM &= (\$9.77\text{M}+\$7.42\text{M})/2 = \$8.595\;\text{million}, \\
\mathrm{TCO}_{\text{CTI}} &= \$0.60\;\text{million/year}.
\end{aligned}
\]

\paragraph{Feasibility result.}
The feasibility constant and frontier are
\[
\kappa_h=\frac{0.60}{8.595}=0.0698,
\qquad
p_{0,h}M>0.0698.
\]
For example, $M=20\%$ requires $p_{0,h}>34.9\%$, while $p_{0,h}=20\%$ requires $M>34.9\%$. Patient-safety consequences may add social value, but their magnitude is outside this financial frontier.

\paragraph{Complementary value indicators.}
The financial frontier captures only part of CTI's potential value in healthcare because cyber incidents can also affect patient safety and human life. Research reports associations between ransomware attacks, care disruption, and increased patient mortality \cite{Ponemon2021, McGlave2023}. These consequences are decision-relevant but remain outside the monetary result reported here.

Healthcare organizations also depend on patient trust in data confidentiality and continuity of care. Because a breach or prolonged outage can impose regulatory costs and erode public confidence, patient retention and satisfaction scores can serve as proxy indicators of value. Survey evidence indicates that up to 66\% of consumers would hesitate to trust a breached company with their data \cite{Semafone2021}. As with patient-safety effects, these trust-related outcomes complement rather than alter the financial frontier.

%% file: Sections/04-Challenges.tex
\section{Measurement Challenges and Methodological Limitations}

Despite the clear value propositions and available methodologies, measuring the ROI of CTI comes with difficulties. These challenges require a deep understanding and a strategic approach to measurement that moves beyond simplistic metrics. Acknowledging these limitations is the first step toward building a more credible and defensible reason for investment.

\subsection{The Prevention Paradox: Quantifying the Value of Non-Events}

The most significant challenge in measuring CTI ROI is the "prevention paradox". The primary goal and greatest success of a CTI program is to prevent security incidents. However, this success is largely invisible, manifesting as an absence of negative events.

It is fundamentally difficult to assign a concrete value to an attack that was avoided or a data breach that never happened. Stakeholders, particularly those in finance and executive leadership, are used to measuring ROI based on tangible gains, such as increased revenue or reduced operational costs. Cybersecurity investments, by contrast, are often justified by the absence of negative outcomes, a concept that can be difficult to translate into a traditional ROI calculation. This paradox often leads to the perception of cybersecurity as a pure cost center, as its benefits are not immediately apparent.

Preventive successes also decay rapidly in institutional memory. Decision-makers often exhibit recency and salience biases, placing greater emphasis on recent and highly visible events than on invisible future risks. A single high-profile breach can override extended periods of effective preventive activity, while a long run of incident-free quarters is quickly normalized as the "expected" state. The CTI team therefore operates under a biasing asymmetry: its work becomes most visible precisely when it fails, and increasingly invisible the longer it succeeds. Because non-events cannot be audited, their value must be inferred: scenario analyses, near-miss retrospectives, red-team simulations, and threat trend timelines provide structured analytical mechanisms for assessing and communicating latent risk. They are not merely narrative devices; they are meant to help decision-makers understand that today's stability is fragile and depends on ongoing effort. In this sense, the prevention paradox reframes ROI as a measure of fragility avoided rather than profit gained \cite{CISA2020}.

\subsection{The Problem of Attribution and Intangibles}

A second major challenge is attribution. In a modern security architecture, a single prevented attack is rarely the result of one control. It is often a combination of CTI providing an early warning, a firewall blocking a malicious IP, an endpoint detection and response (EDR) agent terminating a process, and a user correctly identifying a phishing email. Isolating the specific contribution of the CTI program in this chain of events is extremely difficult.

Furthermore, many of CTI's most critical benefits are intangible. Protecting a company's brand reputation, maintaining customer trust, and boosting employee morale are all significant outcomes of a successful CTI program, but they do not always have a clear, direct monetary value \cite{SANSCtiMetricsChallenges}. While the negative financial impact of losing these assets is evident after a breach, quantifying their value in a preventive context is a complex analytical task.

\subsection{Overcoming Data Overload}

The effectiveness of a CTI program, and consequently its ROI, can be materially degraded by the challenge of intelligence signal dilution. Security teams routinely ingest data from multiple concurrent commercial and open-source feeds while simultaneously processing high volumes of internally generated log and alert telemetry, resulting in an aggregation burden that outpaces available analytical capacity. In the absence of robust aggregation, correlation, and contextualization processes, this signal surplus can precipitate feed abandonment, decision paralysis, or systematic alert desensitization among analysts. Under conditions of sustained noise saturation, operationally relevant intelligence fails to surface, and the program's capacity for anticipatory threat mitigation is substantially diminished.

This problem can be further complicated by a lack of process formalization. Many CTI programs are treated as "academic exercises" rather than operational functions integrated into the security workflow \cite{SANS2025Survey}. They may produce intelligence reports that are interesting but not directly actionable by the SOC or incident response teams.

\subsection{Methodological Approaches to Measurement Challenges}

While these challenges are significant, they can be addressed through a strategic and methodological approach to measurement.

\noindent\emph{Risk quantification.} The most effective way to counter the prevention paradox is to shift the conversation from "what we prevented" to "how much we reduced our risk." Quantitative risk models such as FAIR and the other approaches discussed in Section \ref{sec:quantitativeKPI} are indispensable in this context. Instead of trying to prove a negative (the non-event), the organization can model the Annualized Loss Expectancy (ALE) of a specific threat scenario (e.g., a ransomware attack by a known actor) and then demonstrate how CTI-informed controls reduce the Loss Event Frequency (LEF) or Loss Magnitude (LM). The ROI is then calculated based on this quantifiable reduction in financial risk exposure.

\noindent\emph{Framework-based measurement.} The problem of attribution can be mitigated by using operational frameworks as the unit of measurement. By mapping CTI-driven improvements to the MITRE ATT\&CK framework, it is possible to demonstrate a direct and measurable increase in an organization's defensive coverage against specific adversary TTPs. This shifts the focus from arguing over which single tool deserves credit to demonstrating a holistic improvement in the organization's capability to defend against a relevant threat. This capability improvement is the direct result of the CTI program's work, regardless of the specific tools used for implementation.

Furthermore, the challenge of measuring intangible benefits can be addressed through the use of proxy metrics that carry tangible financial value. Examples include reductions in annual cyber insurance premiums, improved customer retention rates, and the avoidance of public relations or legal costs stemming from security incidents. These proxies serve as concrete indicators of value, translating abstract security gains into measurable business outcomes.

\noindent\emph{Contextualization and automation.} The solution to data overload is generally not less data, but greater context. The value of CTI is unlocked when external threat data is correlated with the organization's internal environment. Implementing a Threat Intelligence Platform (TIP) or a similar aggregation capability is a critical component of this process. These platforms automate the process of ingesting data from multiple feeds, de-duplicating it, and enriching it with internal context, such as asset criticality, user privileges, and vulnerability status. This allows the system to automatically prioritize threats that are not only active in the wild but also pose a direct and immediate danger to the organization's critical assets. The emphasis is therefore on relevance and actionability rather than raw volume.

%% file: Sections/05-Conclusion.tex
\section{Discussion and Conclusion}

Establishing a demonstrable Return on Investment (ROI) for CTI programs has become a critical concern for contemporary cybersecurity: as security budgets face growing financial scrutiny, CTI value must be communicated in terms aligned with business priorities, namely financial risk reduction, operational efficiency, and strategic support. This work has shown that, while significant measurement challenges exist, they can be addressed through a multi-layered methodological approach.

Such an approach cannot rely on a single metric; it combines evidence across four complementary layers, each answering a different stakeholder's question. \emph{Financial quantification} (executives and the board) uses the FAIR model to express CTI's contribution as a reduction in Annualized Loss Expectancy (ALE). \emph{Operational frameworks} (security leadership and auditors) use MITRE ATT\&CK coverage heatmaps as evidence-based proof of improved defensive capability. \emph{Performance KPIs} (SOC managers), such as MTTD, MTTR, and false-positive reduction, capture day-to-day operational impact. Finally, a \emph{qualitative narrative} situates these outcomes within strategic priorities such as brand protection and regulatory compliance. The primary financial result is a two-unknown feasibility frontier: positive ROI requires $p_0M>\mathrm{TCO}_{\mathrm{CTI}}/LM$, where $p_0$ is the organization-specific annual probability of the modeled material event, $M$ is the CTI-attributable mitigation fraction, and $\mathrm{TCO}_{\mathrm{CTI}}$ is the all-in annual incremental investment required to realize that mitigation. Using the scenario loss and all-in cost inputs, the cost-to-loss ratios are 6.87\% for finance and 6.98\% for healthcare. These are not estimated event probabilities or predicted returns; they are the minimum products $p_0M$ that an adopting organization must support with its own evidence. The TIEI-conditioned PERT simulation remains a secondary, assumption-conditional illustration.

\subsection{Implementation Considerations}

Based on this analysis, five considerations support a robust and defensible business case for CTI investment.

\noindent\emph{Early measurement and iteration.} Organizations should not wait for a CTI program to mature before measuring its impact. Baseline metrics for key indicators (e.g., MTTD and false-positive rate) should be captured before new capabilities are deployed, starting with simple, low-effort measures and evolving toward more sophisticated ones. This iterative approach demonstrates continuous improvement \cite{SANSCtiMetricsChallenges}.

\noindent\emph{Automation and integration.} Manual CTI processes are unsustainable and deliver poor ROI given the volume and noise of modern data. A Threat Intelligence Platform, or a comparable capability that automates the collection, processing, and contextualization of intelligence, is essential: automatically correlating external threat data with internal telemetry is the primary mechanism for overcoming data overload and producing prioritized, actionable intelligence \cite{SANSCtiMetricsChallenges}.

\noindent\emph{Complete investment boundary.} The business case must include every incremental downstream cost required to produce the claimed mitigation, even when another team owns the expenditure. Detection engineering, EDR/SIEM/SOAR changes, integration and automation work, control redesign, maintenance, and internal labor therefore belong in $\mathrm{TCO}_{\text{CTI}}$ whenever they are absent from the baseline. Crediting their benefits while excluding their costs creates an incomplete counterfactual and overstates ROI.

\noindent\emph{Audience-specific communication.} Findings should be tailored to each stakeholder \cite{SANSCtiMetricsChallenges}. Because the metrics span technical indicators to strategic outcomes, a single comprehensive report risks overwhelming some readers while under-informing others; multiple targeted summaries, each aligned with its audience's interests and responsibilities, are usually more effective.

\noindent\emph{External benchmarks and peer data.} Internal calculations gain credibility when compared with external evidence. Authoritative industry reports can inform loss magnitude and threat context, while participation in industry ISACs provides peer experience and collaborative intelligence \cite{IBM2024}. Aggregate attack prevalence should not be substituted directly for an organization's probability of a scope-matched material loss event.

\subsection{Limitations of This Study}

The framework proposed here is methodological rather than predictive. Its frontier deliberately leaves two quantities unresolved: the organization-specific probability $p_0$ of the scope-matched material event and the incremental, CTI-attributable mitigation factor $M$. It therefore does not estimate a sector's ROI from public attack prevalence. The loss magnitudes pool two IBM report years, and the all-in program costs are scenario inputs; the reported ranges for $\mathrm{TCO}_{\mathrm{CTI}}/LM$ are deterministic sensitivity bounds, not confidence intervals. The numerical frontiers are valid only if those cost envelopes include every component of $\mathrm{TCO}_{\mathrm{CTI}}$ defined in Equation~\eqref{eq:CTICost}, including downstream enablement. If a cited or budgeted amount covers only core CTI ownership, the organization must identify the omitted components, calculate the complete $\mathrm{TCO}_{\mathrm{CTI}}$, and recompute the frontier. In the secondary simulation, $M$ is represented by a single net factor drawn from a TIEI-conditioned PERT distribution rather than by separately modeled effects on Loss Event Frequency and Loss Magnitude. The PERT anchors and TIEI-to-$M$ coefficients are scenario-analytic calibration assumptions, not empirical estimates, and the selected values of $p_0$ are stress-test inputs, not estimates. This formulation sacrifices point predictions in exchange for preserving event-scope and causal consistency.

\subsection{Limitations of Existing Research and the Need for Independent Evaluation}

A critical observation that emerged during this analysis is the clear dominance of vendor-sponsored research in the field of CTI and its Return on Investment. A substantial portion of the available literature originates from private-sector entities, particularly firms that develop and market Threat Intelligence Platforms. Such contributions offer valuable operational insight and real-world data, but they also introduce a significant potential for conflict of interest.

Vendors have a direct interest in demonstrating the efficacy and business value of CTI solutions, often to justify investment in their proprietary platforms. Studies they sponsor or conduct may therefore emphasize favorable outcomes, selectively present metrics, or frame findings to align with commercial objectives. This funding bias does not necessarily invalidate the data, but it demands careful scrutiny when such studies inform strategic investment decisions. Future work should prioritize independent, academically rigorous studies of CTI effectiveness and value, with open-source intelligence communities and governmental cybersecurity centers contributing to a more diversified research landscape.

\subsection{Directions for Future Research}

The field of CTI is constantly evolving, and so too must the methodologies for measuring its value. Future research should focus on four areas to advance the practice of CTI ROI quantification.

First, the TIEI-to-ALE coupling introduced here should be validated empirically. Longitudinal studies linking measured TIEI scores to observed changes in loss frequency and magnitude would replace the scenario-analytic calibration used in this work with data-driven estimates of CTI's mitigation effect.

Second, the creation of anonymized, industry-wide performance benchmarks for CTI effectiveness metrics (e.g., average MTTD reduction from CTI and typical false-positive reduction rates) would allow organizations to conduct more accurate and meaningful peer comparisons, further strengthening their business cases.

Third, as organizations increasingly rely on AI and machine learning, new threat vectors such as data poisoning and model theft will emerge. Research is needed to develop standardized models for quantifying the ROI of CTI in protecting these systems.

Fourth, as manufacturing and logistics increasingly merge cyber-physical systems with complex vendor ecosystems, research should examine CTI integration across OT/ICS and supply-chain environments. Composite ROI models are needed to capture overlapping dependencies, shared threat intelligence, and coordinated response efficiencies across both domains.

In conclusion, demonstrating the ROI of CTI is complex but achievable. By reporting first the mitigation required to break even, then using maturity-conditioned probability models only as disclosed secondary illustrations, security leaders can communicate the economic decision without presenting uncertain CTI effectiveness as an empirical fact. Combined with operational and qualitative evidence, this approach can reframe CTI from a perceived cost center into a defensible strategic investment.

%% file: Sections/06-Appendix.tex
\appendix
\section{Optional Multi-Anchor Normalization}
\label{app:multi-anchor-normalization}

The default TIEI normalization uses a single performance target, as described in Section~\ref{sec:tiei-scoring-rubric}. In some settings, however, a single target does not adequately represent the intended relationship between performance and operational value. In that case, an organization may use the optional multi-anchor method defined below. The method preserves the TIEI dimensions, weights, geometric aggregation rule, and numerical $[1,100]$ range, but constitutes a distinct calibration regime rather than a plug-compatible replacement. Historical scores must be restated, and maturity cutoffs and any downstream TIEI-to-mitigation calibration must be revalidated before comparison or reuse.

Instead of one target, the organization registers three anchors before the observation period: (i) an \emph{unacceptable boundary}, representing performance that does not meet the minimum operational requirement; (ii) a \emph{reference level}, representing established or expected performance; and (iii) a \emph{stretch level}, representing strong but realistically attainable performance.

For a higher-is-better measure $x$, let $L<R<U$ denote the unacceptable, reference, and stretch anchors, respectively. The normalized score is
\[
N_{\uparrow}(x;L,R,U)=
\begin{cases}
1, & x\le L,\\[4pt]
1+49\dfrac{x-L}{R-L}, & L<x\le R,\\[6pt]
50+50\dfrac{x-R}{U-R}, & R<x<U,\\[4pt]
100, & x\ge U.
\end{cases}
\]
Thus the unacceptable boundary maps to 1, the reference level maps to 50, and the stretch level maps to 100. For example, if validated coverage has anchors $(L,R,U)=(40\%,70\%,90\%)$, an observed coverage of 80\% receives a score of 75.

For a lower-is-better measure $h$, such as MTTD or MTTR, let $L<R<U$ instead denote the stretch, reference, and unacceptable anchors. The direction is reversed:
\[
N_{\downarrow}(h;L,R,U)=
\begin{cases}
100, & h\le L,\\[4pt]
50+50\dfrac{R-h}{R-L}, & L<h<R,\\[6pt]
1+49\dfrac{U-h}{U-R}, & R\le h<U,\\[4pt]
1, & h\ge U.
\end{cases}
\]
For example, if MTTD has anchors $(L,R,U)=(4,12,36)$ hours, an observed MTTD of 8 hours receives a score of 75.

The anchors must not be selected after observing the measurement-period result. They should be chosen using the strongest available evidence, in descending order of preference: regulatory or contractual requirements, established service-level objectives, independent peer benchmarks, risk-model requirements, and a frozen historical baseline. The CTI team should document each anchor's source, rationale, population, observation window, and approval date. Anchors should be approved by relevant stakeholders outside the team being assessed and remain fixed during a longitudinal comparison. If they change, prior scores should be restated where possible; otherwise, the change must be reported as a structural break. Where anchor selection remains uncertain, the reported TIEI should be accompanied by a sensitivity analysis showing how reasonable alternative anchors affect the result.

\section{CTI ROI Indicator Tables}
\label{app:roi-indicator-tables}

\setcounter{table}{0}
\renewcommand{\thetable}{\thesection.\arabic{table}}

\begin{table}[H]
  \caption{ROI Indicators for CTI (Quantitative)}
  \label{tab:roi_indicators_quant}
  \renewcommand{\customtableformatting}{%
    \renewcommand{\arraystretch}{1.0}\footnotesize
  }
  \centering
  \renewcommand{\tabularxcolumn}[1]{>{\centering\arraybackslash}m{#1}}
  \begin{tabularx}{\columnwidth}{|>{\centering\arraybackslash}m{0.26\columnwidth}|X|X|}
    \hline
    \textbf{Metric Category} & \textbf{Indicator} & \textbf{Measurement Method / Tools} \\
    \hline
    \multirow[c]{3}{*}{\begin{tabular}{@{}c@{}}\textbf{Incident} \\ \textbf{Response}\end{tabular}} & Reduction in Mean Time to Detect (MTTD) & SIEM/SOAR Logs; Incident‑Response Reports \\
    \cline{2-3}
    & Reduction in Mean Time to Respond (MTTR) & SIEM/SOAR Logs; Incident‑Response Reports \\
    \cline{2-3}
    & Reduction in Attacker Dwell Time & IR Forensics; Threat‑Hunting Logs \\
    \hline
    \multirow[c]{2}{*}{\begin{tabular}{@{}c@{}}\textbf{Investment} \\ \textbf{Boundary}\end{tabular}} & Core CTI Ownership Cost & Platform, Feed, Staffing, Infrastructure, and Training Budgets \\
    \cline{2-3}
    & Downstream Enablement Cost & Detection-Engineering Work Orders; EDR/SIEM/SOAR Change Records; Internal Labor and Maintenance Allocations \\
    \hline
    \multirow[c]{3}{*}{\textbf{Cost Avoidance}} & Value of Prevented Breaches & FAIR Model; Annual Breach‑Cost Studies (e.g., IBM) \\
    \cline{2-3}
    & Avoided Ransom Payments & CTI on Ransomware Groups; IR Data \\
    \cline{2-3}
    & Reduced Cyber‑Insurance Premiums & Insurance Policy Documents; Risk‑Assessment Reports \\
    \hline
    \multirow[c]{3}{*}{\textbf{Efficiency}} & Reduction in False‑Positive Alerts & SIEM/TIP Platform Metrics \\
    \cline{2-3}
    & Analyst Hours Saved via Automation & Time‑Tracking Studies \\
    \cline{2-3}
    & Increased Threat‑Hunting Success Rate & Threat‑Hunting Reports; Detections per Hypothesis \\
    \hline
    \multirow[c]{2}{*}{\textbf{Brand Protection}} & Takedown of Phishing / Counterfeit Sites & Brand‑Monitoring Platform Metrics \\
    \cline{2-3}
    & Reduction in Customer Fraud Complaints & Customer‑Service Logs; Fraud Reports \\
    \hline
  \end{tabularx}
\end{table}

\begin{table}[H]
  \caption{ROI Indicators for CTI (Qualitative)}
  \label{tab:roi_indicators_qual}
  \footnotesize
  \centering
  \renewcommand{\tabularxcolumn}[1]{>{\centering\arraybackslash}m{#1}}
  \begin{tabularx}{\columnwidth}{|>{\centering\arraybackslash}m{0.26\columnwidth}|X|X|}
    \hline
    \textbf{Metric Category} & \textbf{Indicator} & \textbf{Measurement Method / Tools} \\
    \hline
    \multirow[c]{3}{*}{\textbf{Strategic Value}} & Improved Vulnerability Prioritization & Patch‑Management Reports; CTI‑to‑CVE Correlation \\
    \cline{2-3}
    & Enhanced Executive Decision‑Making & Stakeholder Surveys; Board‑Meeting Feedback \\
    \cline{2-3}
    & Increased Security‑Posture Maturity Score & NIST CSF or CMM Assessments \\
    \hline
    \multirow[c]{1}{*}{\textbf{Brand Protection}} & Improved Brand / Reputation Score & Market Surveys; Net Promoter Score (NPS) \\
    \hline
  \end{tabularx}
\end{table}

\FloatBarrier